\documentstyle[12pt,aasms4]{article}
\input epsf.tex
\begin{document}
\title{Gamma Ray Bursts from Stellar Collisions}
\author{Brad M.S. Hansen \& Chigurupati Murali}
\affil{Canadian Institute for Theoretical Astrophysics \\
University of Toronto, Toronto, ON M5S 3H8, Canada}
\authoremail{hansen@cita.utoronto.ca}
\begin{abstract} We propose that the cosmological gamma ray bursts
 arise from the collapse of neutron stars to black holes triggered
by
 collisions or mergers with main sequence stars.
This scenario represents a  cosmological history
 qualitatively different from most previous theories because
it contains a significant contribution from
 an old stellar population, namely the globular clusters.
  Furthermore, the
gas poor central regions of globular clusters provide an ideal environment for
the generation of the recently confirmed afterglows via the fireball scenario.
Collisions resulting from neutron star birth kicks in close binaries may also
contribute to the overall rate and should lead to associations between 
some gamma ray bursts and supernovae of type~Ib/c.
\end{abstract}

\keywords{stars:neutron--supernovae:general--globular clusters:general--gamma rays:bursts--
Galaxy:evolution}

\section{Introduction}
The detection of X-ray, optical and radio afterglows from
gamma-ray burst events (Costa et al 1997; van Paradijs et al 1997,
Sahu et al 1997; Frail et al 1997) and concommitant redshifts (Metzger
et al 1997) has provided strong support for the cosmological origin
of the phenomenon. The success of the fireball model (Cavallo \& Rees 1978;
Rees \& Meszaros 1992;
Meszaros \& Rees 1993) is largely independent of the origin of the energy,
requiring only an energetic and baryon-poor event confined to small scales,
thereby giving rise to a relativistic and optically thick outflow. These
requirements alone place extreme limits on the nature of the genesis,
for the energy ($\sim 10^{52}$ ergs) and scales ($\sim 10^6 cm$), which
suggest some catastrophic event associated with the birth or death of
compact objects.

Although these are powerful constraints, they leave sufficient leeway
to provide several avenues of investigation, such as binary mergers
involving neutron stars and black holes (Paczynski 1986; Eichler et al 1989;
Paczynski 1991), failed supernovae collapsing to black holes (Woosley 1993),
the collapse of rotating massive stellar cores ('hypernovae') (Paczynski 1997)
or the accretion induced collapse of white dwarfs or neutron stars in binaries
(Usov 1992; Qin et al 1998). All except Usov's scenario are
`prompt' in the sense that the delay between a burst of star formation and the
occurrence of such events is small relative to cosmological timescales. This
holds true even for the binary inspiral scenarios, although they may
possess a long-timescale tail in the delay distribution (Tutukov \& Yungelson 1997;
Portegies Zwart \& Spreeuw 1996). This trend has spawned investigations of the
redshift distribution assuming rates derived from the recently determined
cosmological star formation rates (Totani 1997; Wijers et al 1998).

In this letter we propose that many gamma ray bursts may arise from the collapse of
neutron stars to black holes following stellar
collisions. This scenario differs from the others
mentioned above in that it is dominated by regions of high stellar density and
therefore
contains a significant contribution from globular clusters, an old stellar population.
The collision scenario also results in a potentially significant contribution from
close binaries containing Helium stars, which may explain the recent association
of some gamma ray bursts with type Ib/c supernovae (Galama et al 1998, Wang \& Wheeler 1998).
 In sections~\ref{Scenario}
we describe the manner in which stellar collisions arise, both in globular 
clusters and in younger populations. Section~\ref{rates} provides estimates for
the event rates and section~\ref{Evol} estimates the cosmological evolution
of the gamma ray burst rate in this scenario.

\section{Compact Objects and stellar collisions}
\label{Scenario}

The salient feature of globular clusters which makes them attractive for
our purpouses is the greatly enhanced probability of compact object
recycling because of the large central densities and high probability
of stellar interaction. The strongest empirical evidence of this is the
high rate of occurrence of recycled millisecond pulsars  in globular
clusters, relative to the galactic disk (e.g. Taylor, Manchester \& Lyne 1993), despite
the greater difficulty of detection.
This is believed to result from exchanges of neutron stars into binaries
or close tidal encounters between a neutron star and a main sequence star (e.g.
Hut et al 1992). The formation of these binaries then leads to spin-up of
the neutron star due to subsequent mass transfer from the newly acquired companion.

The existence of millisecond pulsars is empirical evidence that such events occur at
significant rates. 
 Theoretical investigations of the cross-sections for
these processes and their applications to typical cluster cores (Davies \&
Benz 1995; Sigurdsson \& Phinney 1995; Davies \& Hansen 1998) suggest that the production of
these binaries also results in merged systems (`smothered neutron stars') at rates equal to or greater than
the binary production rate. The fate of these systems is much less certain.
Depending on the angular momentum of the encounter, the neutron star may spiral
to the center of the collision remnant or it may disrupt the star, forming a
thick accretion torus (Davies et al 1992; Rasio  1993).

The limiting accretion rate onto a neutron star is considerably larger
than the Eddington rate because of neutrino losses (Zeldovich, Ivanova \& Nadezhin 1972;
Chevalier 1993;
Fryer, Benz \& Herant 1996), so that the formation of such merged systems is likely
to  lead to rapid accretion of the disrupted companion material on approximately
free-fall times for quasi-spherical remnants and global viscous transport times for disks
 (Chevalier 1993). The rapid accretion of most of
the remnant mass will lead to a collapse to a black hole. The final stages described here
bear some resemblance to the gamma ray burst mechanism described by Woosley (1993) (where the accreted material
was from supernova fallback). 

Thus, we suggest that the production of stellar mass black holes is an inevitable 
consequence of the production of millisecond pulsars in globular clusters (the essential
difference in the two outcomes is the rate of accretion, the latter case being regulated
by the relatively miserly donation rate of the binary companion).
The resultant rapid energy release on small scales is similar to several gamma ray burst
scenarios. As we shall see, the hypothesis that such events do give rise to gamma
ray bursts yields several interesting and potentially testable consequences.
In particular, it represents a
formation history qualitatively different from most other proposed mechanisms, as it is not
tied directly to the star formation rate.

Younger stellar populations in galactic disks have much lower densities and thus the 
three body encounter rate is negligible. Nevertheless, many supernova progenitors are
found in binaries and the evidence is mounting that neutron stars acquire significant
velocities at birth (e.g. Lyne \& Lorimer 1994; Cordes, Romani \& Lundgren 1993). 
If the velocity is directed at a binary companion, the resulting stellar collision
may produce a similar burst event. Although such an occurrence is a priori unlikely, the
gamma ray bursts are themselves rare events, and we shall show that this rate may be
significant.

\section{Rates}
\label{rates}
If we adopt the position that such stellar collapses occur and give rise to 
gamma-ray bursts, we are still left with the requirement that they produce a
significant rate. Canonical rates for cosmological gamma-ray burst production are often based
on the estimated rate of binary neutron star mergers (Phinney 1991; Narayan, Piran
\& Shemi 1991), $\sim$1~event~Myr$^{-1}$ per galaxy, and in agreement with
fits to the log~N-log~P relation for nonevolving cosmological rates (Fenimore \& Bloom 1995).
However, recent analyses of the latter type, using rates based on  cosmological star formation histories
(as befits the prompt burst production scenarios), have found rates lower than this
by factors of $\sim 100$ (Wijers et al 1998) to $\sim 1000$ (Totani 1997). Thus any
rate in the range $10^{-3}-1$~Myr$^{-1}$ per galaxy is of significance.

The rate of stellar collisions  increases dramatically with central cluster
density, and thus the total rate will be dominated by those clusters which
are undergoing the `binary burning' phase thought to provide support against
core collapse (e.g. Goodman 1988). Indeed, many millisecond pulsars
are found in post core-collapse clusters such
as M15 or in the highest density non-collapsed clusters such as 47 Tuc.

Burst rate estimates based on theoretical cross-sections and globular cluster evolution are model-dependent
but we may obtain robust empirical estimates with the modest assumption that the burst rate is
comparable to or larger than the rate of millisecond pulsar production (see section~\ref{Scenario}).
The pulsar birthrate may be estimated by considering the characteristic spin-down ages of the
individual pulsars and the selection effects involved in the detections. Phinney (1996) estimates
 density dependent production rates of $10^{-8}$ yr$^{-1}$, $2 \times 10^{-9}$ yr$^{-1}$ and
$5 \times 10^{-10}$ yr$^{-1}$ in individual clusters of central luminosity density $10^6 L_{\odot} pc^{-3}$,
$10^5 L_{\odot} pc^{-3}$ and $10^4 L_{\odot} pc^{-3}$ respectively. Note that these refer only to
currently observable pulsars and are thus almost certainly lower limits, given the steepness of
the disk pulsar luminosity function and the large distances to globular clusters. Phinney estimates
the true rate could be $\sim 100$ larger. Assuming $\sim 20\%$ of the globular clusters in the galaxy
have sufficiently high central densities (Djorgovski \& King 1986; Chernoff \& Weinberg 1990),
 this means the  event rate for our
galaxy can lie anywhere in the range $10^{-2}-10$~Myr$^{-1}$. Hence, relative to the new lower rates
inferred by Totani \& Wijers et al, the globular clusters are almost certainly important, and may
even be important for the original non-evolving estimates.

The contribution from the galactic disk will be dominated by close binaries. The supernovae which
occur in such systems appear as type Ib/c at a rate $\sim 10^{-3} yr^{-1}$ and studies of 
low mass X-ray binary formation suggest pre-supernova separations of $\sim 10-25 R_{\odot}$ (e.g. Kalogera \& Webbink 1998).
Assuming randomly directed kicks from the supernova, the collision rate (for companions of approximately
solar mass) is $\sim 1 Myr^{-1}$ in our galaxy. Hence, this pathway may also be important.


\section{Cosmological Evolution}
\label{Evol}

A gamma-ray burst history dominated by globular cluster production will
provide a cosmological history qualitatively different from the currently fashionable one
based on star formation rate. Although the epoch of globular cluster formation is
somewhat theory dependent (see Fall \& Rees 1988 for a review), globular cluster
ages (Chaboyer, Demarque \& Sarajedini 1996) suggest that a significant fraction of the galactic system must have existed since
high redshifts. 

To illustrate the cosmological history of globular cluster burst production, we have calculated
the evolution of a globular cluster population assumed to have formed in a single burst 15 Gyr ago.
Following Murali \& Weinberg (1997) we assume 250 clusters with an initial mass distribution
$dN/dM \propto M^{-1.75}$ in mass range $10^5-10^6 M_{\odot}$ and spatial distribution $\rho \propto R^{-2.75}$ in
a Milky-Way like galaxy. Individual clusters are evolved using the Fokker-Planck models of
Murali \& Weinberg (1997) on circular orbits in the spheroidal potential. Extra heating 
due to eccentric orbits and the disk potential do not change the evolution qualitatively.
 Figure~\ref{murali} shows the evolution of total integrated
core mass in regions of density above the threshold (taken to be $10^3 M_{\odot} pc^{-3}$)
 in the cluster system as a function
of time. The initial rise comes from low mass clusters at small radii (which are the dominant
population and which have the shortest
evolutionary timescale). 
Eventually the integrated core mass drops again as the low mass clusters
evaporate completely.
Although a galaxy globular cluster system is influenced by environment (e.g. cluster
cD galaxies such as M87 have much larger globular cluster systems), the number of
clusters per galaxy scales roughly with luminosity within a given Hubble type
 (Harris \& Racine 1979; VanDenBerg 1984; Harris 1991). Thus, for the illustrative
discussion below, we will adopt our above
rate as a function of time per galaxy for all types (since spirals and ellipticals have similar
numbers of clusters on average) and simply normalise to galactic luminosity 
densities.


Figure~\ref{NPplot} shows the $\log N - \log P$ relation from the BATSE 3B catalogue
 (Meegan et al 1996) for the
1024ms channel. The fit is for peak fluxes $P > P_{th} = 0.4$~photons~cm$^{-2}$~s$^{-1}$
and assuming a single, standard candle luminosity and the cosmological history determined above.
The threshold redshifts for currently favoured open and $\Lambda$ models (see Ostriker \&
Steinhardt 1995) lie in the range 2.5-5 depending on the assumed spectral index, taken
to be $\alpha=1.1\pm 0.3$ (Mallozzi, Pendleton \& Paciesas 1996). The derived isotropic standard
candle luminosities are $\sim 5 \times 10^{52}$ ergs. The required rate, using the estimated
globular cluster comoving space density $\sim 10 h^3$Mpc$^{-3}$ (Phinney 1991), is then only $ 3 \times
10^{-9} yr^{-1}$ per globular cluster. When this average is only taken over those undergoing
core collapse, then the inferred rates are similar to the low end estimated in section~\ref{rates}.
 The point of this simple
comparison with the data is to demonstrate that globular cluster scenarios can naturally
incorporate high redshift events and any significant contribution from globular clusters
is likely to move the average redshifts outwards. In fact, Totani (1997)  finds that simple 
comparisons like this one but using rates based on empirical star formation laws are improved
by 
an additional contribution from high redshifts, which he attributes to an early epoch of
elliptical galaxy formation. Globular cluster contributions can serve this purpouse equally
well. The true rate is likely to be a combination of this one and one that does follow the
star formation rate if the contribution from close Helium star binaries is important.
Indeed, despite the very different conceptual histories, the globular cluster and star formation
rates display similar redshift dependances, since both increase monotonically for several Gyr and
then fall away at later times.

\section{Consequences and Conclusions}
\label{obs}

The gamma-ray burst scenario described here resembles Woosley's mechanism in terms
of physical processes, but contains an important difference with respect to most 
 other cosmological gamma-ray burst models, namely a significant contribution
from old stellar populations. As such, afterglow searches should be able to distinguish
between this and other models by identifying host galaxies. Most of the currently
accepted burst models suggest that afterglows should be associated with galaxies undergoing vigorous star
formation, while our scenario will produce afterglows associated with galaxies across the
Hubble sequence.

A particularly attractive feature of the globular cluster contribution is that it provides
a naturally gas-poor environment for the fireball to propagate. Although the high
stellar density implies that $\sim 10^3 M_{\odot}$ of material should be deposited in the
cluster cores from giant star mass loss between galactic disk passages, observations indicate
that globular cluster cores are remarkably gas poor (see Roberts 1988; Knapp et al 1996 for reviews).
 Typical upper limits are $\sim 0.1 - 1 M_{\odot}$
in total. IRAS measurements (Origlia et al 1996) infer warm dust masses $\sim 10^{-8}-10^{-6} M_{\odot}$ (implying gas masses $\sim 100$ times larger). Furthermore, they present evidence that this
is concentrated in unresolved circumstellar envelopes, so that average intracluster values are
even lower. Explanations range from ram pressure stripping, photoionisation (see above reviews)
to evacuation due to the accumulated wind from millisecond pulsars (Spergel 1991).
Thus, ambient gas masses $\sim 10^{-6} M_{\odot}$ on scales $\sim 0.1-1$pc are generic,
making globular cluster cores an ideal environment for the creation of gamma ray bursts
and their afterglows from fireball models.
 
Gamma ray bursts associated with post-supernova collisions in binaries may also
be identifiable. While not wishing to adopt any particular model for burst evolution,
we note that gamma ray bursts in these events should be associated with type Ib/c
supernovae (which provide the supernova kick). Of course, the interaction of the
two related explosions is likely to cause deviations from standard light curves.

As noted by Wijers et al (1998), a significant contribution from high redshift sources is
helpful in explaining the lack of observed host galaxies for some afterglows. The 
large redshift weighting in this scenario will make constraints from Lyman limit 
observations (Bloom et al 1997) important.

In addition to the possible link to gamma-ray bursts, the production of black holes
described in this paper is interesting in it's own right. The presence of stellar
mass black holes resulting from very massive stars (Larson 1984; Kulkarni, Hut \& McMillan 1993;
Sigurdsson \& Hernquist 1993) has profound consequences for globular cluster structure and
evolution by virtue of their larger masses and greater central concentration. The black hole
population we describe differs from the prior hypotheses because it is not a product of
initial conditions, but rather a product of cluster evolution and stellar interaction. The
final population will be determined by the competition between the production and 
depletion via mass segregation and dynamical ejections. Elucidation of this process may
allow us to place constraints on this scenario from the apparent paucity of black hole
X-ray binaries in globular clusters (in't~Zand et al 1998)

In conclusion, we have proposed that the processes which produce  millisecond pulsars in globular clusters
 also produce  a significant population of stellar mass black holes and
may provide the source of gamma ray bursts. In particular, the propagation of a fireball
is aided by the evacuated nature of cluster cores, possibly by the millisecond pulsar winds
themselves. The fact that these bursts arise from an old stellar population suggests a significant
contribution from high redshifts, an assertion testable with detailed monitoring of the optical
afterglows. The equivalent mechanism in younger stellar populations should lead to an
association of some bursts with type Ib/c supernovae.

The authors would like to thank Phil Armitage, Fred Rasio and Vicky Kalogera for illuminating discussions and
for noting the importance of the close binary contribution. Brad Hansen would also like to
thank the Aspen Center for Physics for their hospitality while some of this paper was written.


\figcaption[murali.ps]{The evolution of the integrated mass in regions above a critical
density $10^3 M_{\odot} pc^{-3}$ for an entire galactic globular cluster system. The
rate of stellar collisions should evolve in a similar fashion.
\label{murali}}

\figcaption[NP.ps]{ The solid histogram is the number-peak flux relation from the BATSE~3B catalogue
using the 1024 ms peak flux. 
The dotted line is for $\Omega_0=0.3$,$\Omega_{\Lambda}=0.7$ and h=0.6. The dashed
line is for the same but $\Omega_{\Lambda}=0$. These curves assume a spectral index
$\alpha=1.1$. 
\label{NPplot}}


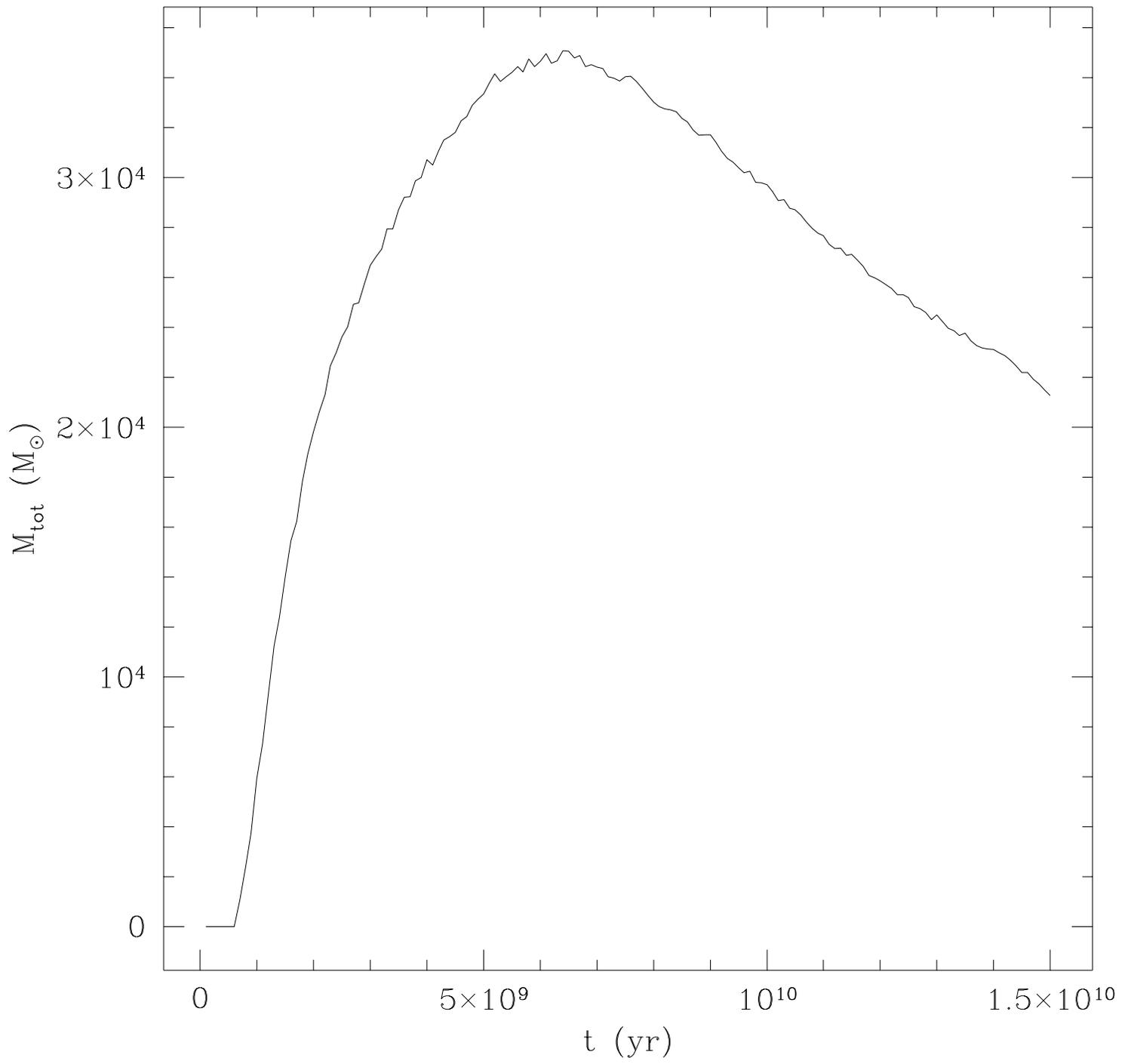
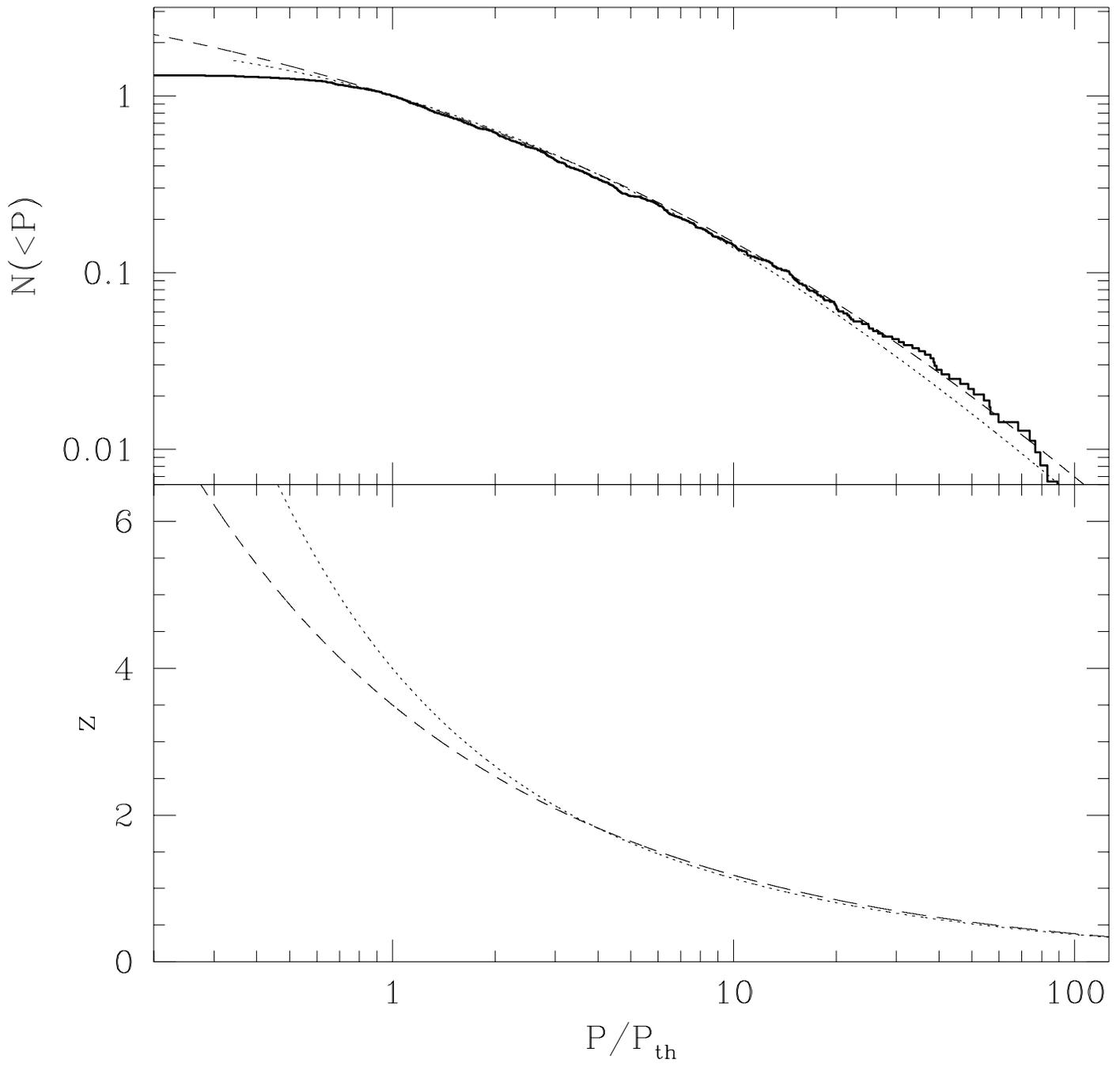
\end{document}